\documentstyle[12pt]{article}

\font\tenmsb=msbm10 scaled\magstep 1
\font\sevenmsb=msbm7 scaled \magstep 1
\font\faivemsb=msbm5 scaled \magstep 1
\newfam\msbfam
\textfont\msbfam=\tenmsb
\scriptfont\msbfam=\sevenmsb
\scriptscriptfont\msbfam=\faivemsb\def\Bbb#1{{\fam\msbfam #1}}

\font\tengothic=eufm10 scaled\magstep 1
\font\sevengothic=eufm7 scaled\magstep 1
\newfam\gothicfam
\textfont\gothicfam=\tengothic
\scriptfont\gothicfam=\sevengothic

\newcommand{\be}{\begin{equation}}
\newcommand{\ee}{\end{equation}}
\newcommand{\dlt}{\delta}
\newcommand{\Dlt}{\Delta}
\newcommand{\ra}{\rightarrow}
\newcommand{\vp}{\varphi}

\newcommand{\al}{\alpha}
\newcommand{\prt}{\partial}

\newcommand{\Om}{\Omega}
\newcommand{\om}{\omega}

\newcommand{\lbd}{\lambda}
\newcommand{\gm}{\gamma}
\newcommand{\sgm}{\sigma}
\newcommand{\ep}{\varepsilon}
\newcommand{\cH}{{\cal H}}
\newcommand{\cD}{{\cal D}}
\newcommand{\cL}{{\cal L}}
\newcommand{\cA}{{\cal A}}
\newcommand{\bk}{{\bf k}}
\newcommand{\br}{{\bf r}}

\newcommand{\bj}{{\bf j}}
\newcommand{\bE}{{\bf E}}
\newcommand{\bH}{{\bf H}}
\newcommand{\bA}{{\bf A}}
\newcommand{\bS}{{\bf S}}

\begin{document}

\begin{center}
{\Large{\bf Entanglement Production under Collective Radiation} \\ [5mm]
V.I. Yukalov} \\ [3mm]
{\it
Bogolubov Laboratory of Theoretical Physics, \\
Joint Institute for Nuclear Research, Dubna 141980, Russia}

\end{center}

\vskip 2cm

\begin{abstract}

The relation is studied between the entanglement production and
collective radiation by an ensemble of atoms. Entanglement production
is quantified by means of a general measure introduced earlier by the
author. Primary emphasis is placed on the entanglement generated by 
pseudospin density matrices. The problem of collective atomic radiation 
can be described by the pseudospin evolution equations. These equations 
define the evolutional entanglement generated by the related density 
matrices. Under conditions of superradiant emission, the entanglement 
production exhibits sharp peaks at the delay time, where the intensity 
of radiation is maximal. The possibility of regulating the occurrence of 
such peaks by punctuated superradiance is discussed, which suggests the 
feasibility of {\it punctuated entanglement production}.

\end{abstract}

\newpage

\section{Introduction}

The concept of entanglement is at the heart of modern ideas on quantum
information processing and quantum computing [1--5]. There exist different
suggestions for manipulating quantum entanglement, in particular, by employing
spin systems or considering atomic systems interacting with electromagnetic
fields [6]. The aim of the present paper is to analyze the relation between
entanglement and an ensemble of radiating atoms. A specific feature of this
analysis is the consideration of a large ensemble of many atoms under
conditions of well developed collective effects and coherent radiation.

First of all, it is necessary to emphasize that, generally, there are two
kinds of entanglement, the entanglement of a quantum state and the entanglement
produced by a quantum operation. And one has to distinguish between these two
principally different notions.

For pure quantum states, represented by wave functions, the concept of
entanglement is straightforward. A pure state of a complex system is termed
entangled if and only if it cannot be represented as a tensor product of wave
functions pertaining to different parts of the given system.

For mixed states, represented by statistical operators, the notion of
entanglement is essentially more complicated. This is because functions
and operators are fundamentally different mathematical quantities. Despite
of this, the name of a state is also commonly applied to statistical, or
density, operators. One decides whether a statistical operator (a mixed state)
is entangled or not according to the formal structure of the given operator.
An operator $\hat\rho$ is called separable if and only if it can be written
as a convex combination of product states:
$$
\hat\rho = \sum_\nu \lbd_\nu \otimes_j \rho^{(j)}_\nu \; ; \qquad
\lbd_\nu\geq 0 \; , \; \;  \sum_\nu \lbd_\nu = 1 \; .
$$
When $\hat\rho$ is not separable it is named entangled. However, from the
mathematical point of view, the properties of an operator are correctly
defined not according to its formal appearance but respectively to its
action on a given set of functions.

In order to better understand the difference between the structure of an
operator, be it a statistical operator or not, and its action, it is useful
to distract ourselves, for a while, from physical applications and to clear
cut the point in mathematical terms. Then, for a given operator, one could
define two types of entanglement. One is the entanglement {\it of an operator}
and another one is the entanglement {\it produced by an operator}. In the
first case, one treats an operator as being entangled or not according to
its formal structure of being nonseparable or separable. In the second case,
one should characterize an operator as {\it entangling} or not on the basis
of its action over a set of disentangled functions. Pay attention that
entangled and entangling are quite different characteristics.

In physical applications, one usually discusses the first type of
entanglement for a particular class of operators, that is, for statistical,
or density, operators. The entanglement of a statistical operator is commonly
termed the entanglement of state. A measure for this type of entanglement
can be well defined only for pure bipartite systems. No general measure of
this type is known for multipartite or mixed systems. It is worth stressing
that, whatever entanglement of a state one is talking about, be it the
entanglement of formation, entanglement cost, relative entropy of entanglement,
or entanglement of distillation, all of them pertain to the same type
characterizing the entanglement of an operator.

The entanglement {\it produced by an operator} is a qualitatively different
notion. To my understanding, only this type of produced entanglement, or
entanglement production, can be correctly defined in mathematical terms.
And it is just the entanglement production that is considered in the present
paper. A general measure of entanglement production, realized by an arbitrary
operator and for systems of any nature, be it multipartite or mixed systems,
has recently been introduced [7,8]. This measure is studied here for
characterizing entanglement produced in the process of collective atomic
radiation.

\section{Entanglement Production}

First, we need to describe a general way of quantifying entanglement production.
Consider a composite system consisting of parts enumerated by the index
$i\in\{ i\}$ pertaining to a label manifold $\{ i\}$. For the standard case of
a discrete manifold, one has $i=1,2,\ldots$ In general, the label manifold can
be continuous. Let each part be characterized by a single-partite Hilbert space
$$
\cH_i \equiv \overline\cL \{ |n_i>\} \; ,
$$
being a closed linear envelope of a single-partite basis $\{|n_i>\}$. The
composite Hilbert space
\be
\label{1}
\cH \equiv \otimes_i \cH_i
\ee
is the tensor product, which is a closed linear envelope
$$
\cH \equiv \overline\cL \{ |n >\}
$$
of a multipartite basis $\{|n>\}$ consisting of the vectors
\be
\label{2}
|n>\; \equiv \otimes_i \; |n_i> \; .
\ee
For a continuous label manifold $\{ i\}$ the products (1) and (2) are the
continuous tensor products, introduced by von Neumann [9] and employed for
particular cases in Refs. [10,11]. In each space $\cH_i$, for functions $\vp_i$
and $\vp_i'$, a scalar product $(\vp_i,\vp_i')$ is defined. The norm, generated
by this product, is
\be
\label{3}
||\vp_i||_{\cH_i}\; \equiv \sqrt{(\vp_i,\vp_i)} \; ,
\ee
which is termed the vector norm.

The composite space (1) can be decomposed into two sets of functions. One sort
of functions $f\in\cH$ is the class of {\it factor functions}
\be
\label{4}
f \equiv \otimes_i \vp_i \qquad (\vp_i\in\cH_i) \; ,
\ee
which are called {\it disentangled functions}. All such functions form the
{\it disentangled set}
\be
\label{5}
\cD \equiv \{ \otimes_i\vp_i|\; \vp_i\in \cH_i\} \; .
\ee
The remaining vectors of $\cH$ constitute the complement $\cH\setminus\cD$ 
whose elements cannot be represented as tensor products of $\vp_i\in\cH_i$. 
Hence, the set $\cH\setminus\cD$ can be named the {\it entangled set}. 
By construction, 
$$
\cH = \cD \cup \cH \setminus \cD \; .
$$
The set $\cD$ is not necessarily a subspace of $\cH$. Nevertheless, on the
basis of the norms (3), it is straightforward to define the restricted vector
norm
\be
\label{6}
||f||_\cD \equiv \prod_i ||\vp_i||_{\cH_i}
\ee
over the set $\cD\subset\cH$.

Let an operator $A$ be given on $\cH$. For instance, this can be an operator
of an observable quantity, or an operator representing a measurement or any
other operation on $\cH$. Assume that the operator $A$ is bounded, possessing
the norm $||A||_\cH$. We may introduce the restricted norm
\be
\label{7}
||A||_\cD \equiv \sup_{f\in\cD}\; \frac{||Af||_\cD}{||f||_\cD} \; ,
\ee
interpreted as the supremum of the expression corresponding to the standard
definition of a norm over an Hilbert space, restricted to the subset $\cD$
of disentangled functions. The operator norm $||A||$, generated by the vector
norm $||Af||$, is also called the Hermitian or spectral norm. The norm (7)
can be written as
\be
\label{8}
||A||_\cD = \sup_{f,f'\in\cD} \; \frac{|(f,Af')|}{||f||_\cD ||f'||_\cD} \; ,
\ee
where $f,f'\neq 0$. For a self-adjoint operator, this can be reduced to
\be
\label{9}
||A||_\cD = \sup_{f\in\cD} \; \frac{|(f,Af)|}{||f||^2_\cD} \; ,
\qquad (A^+=A) \; ,
\ee
where again $f\neq 0$.

Any factor function $f\in\cD$ can be written as an expansion
$$
f=\otimes_i\; \sum_{n_i} a_{n_i}|n_i> \; ,
$$
which can be represented as
$$
f=\sum_n c_n|n> \qquad (n=\{n_i\}) \; ,
$$
with
$$
c_n =\prod_i a_{n_i} \; , \qquad |n>\; \equiv \otimes_i\; |n_i> \; .
$$
Therefore, norm (9) takes the form
\be
\label{10}
||A||_\cD = \sup_{\{ c_n\} } \;
\frac{|\sum_{mn} c_m^* c_n<m|A|n>|}{\sum_n |c_n|^2} \; .
\ee
The latter, keeping in mind an orthonormalized basis, such that $<m|n>=\dlt_{mn}$,
can be simplified to
\be
\label{11}
||A||_\cD = \sup_n |<n|A|n>| \; .
\ee
This representation is the most convenient for practical calculations.

For each operator $A$ on $\cH$, one may put into correspondence a nonentangling
operator $A^\otimes$ having the structure of a tensor product $\otimes_i A_i$
of single-partite operators
\be
\label{12}
A_i \equiv {\rm Tr}_{\{ \cH_{j\neq i}\} } \; A \; .
\ee
To preserve the scale-invariant form of $A^\otimes$, we require the validity of
the normalization condition
\be
\label{13}
{\rm Tr}_\cH \; A^\otimes = {\rm Tr}_\cH \; A \; ,
\ee
in which
$$
{\rm Tr}_\cH\; \otimes_i A_i = \prod_i {\rm Tr}_{\cH_i} \; A_i \; .
$$
The amount of entanglement produced by an operator $A$ is characterized [7,8] by
the measure of entanglement production, which compares the actions of $A$
and $A^\otimes$ on the disentangled set $\cD$. The nonentangling counterpart of
$A$, taking account of condition (13), reads
\be
\label{14}
A^\otimes \equiv \frac{{\rm Tr}_\cH A}{{\rm Tr}_\cH \otimes_i A_i}\;
\otimes_i A_i \; .
\ee
The {\it measure of entanglement production} by an operator $A$ is
\be
\label{15}
\ep(A) \equiv \log \; \frac{||A||_\cD}{||A^\otimes||_\cD} \; .
\ee
Here the norm $||A||_\cD$ is given by Eq. (11) and the norm $||A^\otimes||_\cD$
can be calculated as
$$
||A^\otimes||_\cD = \frac{{\rm Tr}_\cH A}{{\rm Tr}_\cH \otimes_i A_i}\;
\prod_i ||A_i||_{\cH_i} = ({\rm Tr}_\cH A) \prod_i
\frac{||A_i||_{\cH_i}}{{\rm Tr}_{\cH_i} A_i} \; .
$$
Measure (15) will be used in what follows for particular physical applications.
By construction, this measure satisfies all properties typical of any measure
[7,8].

Several remarks are in order. When one is interested in the entangling properties
of a family $\cA\equiv\{ A\}$ of operators, one may calculate measure (15) for
each of them and then find the maximal value
$$
\ep(\cA) \equiv \sup_{A\in\cA}\; \ep(A)
$$
quantifying the entanglement production by the family $\cA$. In particular,
$\cA$ can represent an algebra of observables.

By replacing in the above expressions the disentangled set $\cD$ by the
entangled set $\cH\setminus\cD$, we can introduce the measure of disentanglement
production
$$
\dlt(A) \equiv \log\;
\frac{||A||_{\cH\setminus\cD}}{||A^\otimes||_{\cH\setminus\cD}} \; ,
$$
describing the disentangling properties of an operator $A$.

The restricted norm (11) does not necessarily coincide with the standard norm,
so that, generally speaking,
$$
||A||_\cD\neq ||A||_\cH \; ,
$$
though in many cases they can be identical. An example, when these norms are
different, can be composed as follows. Let $\vp\in\cH$, with $||\vp||=1$, can
be written as
$$
\vp = \sum_n c_n|n> \qquad (n\equiv\{ n_i\}) \; .
$$
Define the operator
$$
A\equiv |\vp><\vp| \; ,
$$
for which $||A||_\cH=1$. At the same time, the restricted norm (11) gives
$$
||A||_\cD =\sup_n | c_n|^2 \; .
$$
Assuming the case, when $\sum_n|c_n|^2=1$ and at least two of $c_n$ are
nonzero, one has $\sup_n|c_n|^2<1$. Thence
$$
||A||_\cD < ||A||_\cH = 1 \; .
$$

In general, the restricted norm $||A||_\cD$ can be expressed through the
standard norm $||A||_\cH$ as follows. Introduce the projector $P_\cD$, which
projects the composite space $\cH$ onto the disentangled set $\cD$, so that
$P_\cD\cH=\cD$. Then, we may write
$$
||A||_\cD \equiv ||P_\cD A P_\cD||_\cH \; .
$$
From here, $||A||_\cD\leq||A||_\cH$. The projector $P_\cD$, however, is
nonlinear, hence $P_\cD AP_\cD$ is a nonlinear operator, even if $A$ is 
linear. Note that the operator norm is usually defined for linear operators.

For pure bipartite systems, one measures entanglement by means of the reduced
von Neumann entropy
$$
S_N \equiv -{\rm Tr}_{\cH_i}\; \hat\rho_i \log\hat\rho_i \; ,
$$
in which
$$
\hat\rho_i \equiv {\rm Tr}_{\cH_{j\neq i} }\; \hat\rho \; ,
$$
and $\hat\rho$ is a statistical operator of the bipartite system. Then
$$
S_N = - \sum_n |c_n|^2 \log|c_n|^2 \; ,
$$
where $|c_n|$ is a Schmidt weight. The entropy $S_N$ is assumed to quantify
the entanglement of a bipartite state. This should not be confused with the
entanglement produced by the statistical operator $\hat\rho$, which is
quantified by measure (15) yielding
$$
\ep(\hat\rho) = - \log \sup_n|c_n|^2 \; .
$$
As is evident, these measures are, in general, different, coinciding only
for the maximally entangled states, for which $|c_n|=const$.

To stress once again that entanglement production by an operator and
entanglement of an operator are rather different notions, let us consider
the Werner operator [12], which is defined for bipartite systems of the
composite space $\cH=\cH_1\otimes\cH_2$, with the single-partite space
dimensionality $d\equiv{\rm dim}\cH_i$. The Werner operator is
$$
\hat W_\al \equiv \frac{1}{d^3-d} \left [ (d-\al)\hat 1 +
(d\al - 1) \hat\sgm \right ] \; ,
$$
where $\al$ is a real number, $\hat 1$ is a unity operator
$$
\hat 1 \equiv \sum_{i,j=1}^d \; |n_i n_j><n_j n_i| \; ,
$$
and $\hat\sgm$ is a flip operator
$$
\hat\sgm \equiv \sum_{i,j=1}^d \; |n_i n_j><n_i n_j| \; .
$$
The name of the latter operator comes from the property
$$
\hat\sgm (\vp_1\otimes\vp_2) = \vp_2 \otimes \vp_1 \; .
$$
Also, one has
$$
{\rm Tr}_\cH \hat W_\al =1 \; , \qquad \al= {\rm Tr}_\cH \hat W_\al
\hat \sgm \; .
$$
The positive partial transpose criterion [13,14] tells us that $\hat W_\al$
is separable if and only if $\al\geq 0$. Because of this, the entanglement
of the Werner operator for $\al\geq 0$ must be zero. However, this operator
is {\it entangling} for all $\al\neq 1/2$, which means that it transforms
nonentangled functions $f\in\cD$ from the disentangled set $\cD$, which
are the factor functions of the form
$$
f = \sum_{i=1}^d a_i|n_i> \otimes \sum_{j=1}^d b_j|n_j> \; ,
$$
into entangled functions $\hat W_\al f\in\cH\setminus\cD$. Therefore, the
measure of entanglement production $\ep(\hat W_\al)$ must be nonzero. For
instance, in the case of two-dimensional single-partite spaces, when $d=2$,
we have
$$
\hat W_\al = \frac{1}{6}\; \left [ (2-\al)\hat 1 + (2\al -1)\hat\sgm
\right ] \; .
$$
From here, we find
$$
\hat W_\al^\otimes = \frac{1}{4}\; \hat 1 \; , \qquad
||\hat W_\al^\otimes||_\cD = \frac{1}{4} \; , \qquad
||\hat W_\al||_\cD = \frac{1}{6}\; \sup\{ 2-\al,1+\al \} \; .
$$
Then, measure (15) becomes
$$
\ep(\hat W_\al) = \log\left ( 1 + \frac{1}{3}\left | 2\al -1 \right |
\right ) \; ,
$$
which is nonzero for all $\al\neq 1/2$. Another example of an operator that
is separable, but at the same time entangling, can be found in Ref. [8].

\section{Spin Entanglement}

The process of radiation by resonant atomic systems can be formulated in
terms of the evolution equations for spin operators. The latter are often
called pseudospin operators in order to emphasize that they do not
correspond to actual spins but are just mathematical structures possessing
the properties of spin operators. For resonant, effectively two-level, atoms,
such operators correspond to spin-1/2 operators, which is implied in what
follows.

It is convenient to deal with the spin operators $S_j^\al\equiv S^\al(\br_j)$
being either the ladder operators $S_j^\pm$ or the $z$-component operator
$S_j^z$. The former are used for describing dipole transitions, while the 
latter represents the population difference. These operators possess the
properties
$$
S_i^\pm S_i^z = \mp \; \frac{1}{2}\; S_i^\pm \; , \qquad
S_i^z S_i^\pm = \pm \; \frac{1}{2}\; S_i^\pm \; ,
$$
$$
S_i^+ S_i^- = \frac{1}{2} + S_i^z \; , \qquad S_i^- S_i^+ =
\frac{1}{2}- S_i^z \; , \qquad S_i^\pm S_i^\pm = 0 \; , \qquad
S_i^z S_i^z = \frac{1}{4}
$$
and satisfy the commutation relations
$$
[ S_i^+,\; S_j^-] = 2\dlt_{ij} S_i^z \; , \qquad [S_i^\pm,\; S_j^z] =
\mp \dlt_{ij} S_i^\pm \; .
$$

Let us introduce the spin density matrices [15] of two types. One type is
$$
R_n = \left [ R_n(i_1\ldots i_n,j_1\ldots j_n)\right ] \; ,
$$
with the elements
\be
\label{16}
R_n(i_1\ldots i_n,j_1\ldots j_n) \equiv \;
< S_{j_n}^+\ldots S_{j_1}^+ S_{i_1}^-\ldots S_{i_n}^- > \; .
\ee
And another type is
$$
Z_n = \left [ Z_n(i_1\ldots i_n,j_1\ldots j_n) \right ] \; ,
$$
with the elements
\be
\label{17}
Z_n(i_1\ldots i_n,j_1\ldots j_n) \equiv \;
< S_{j_n}^z\ldots S_{j_1}^z S_{i_1}^z\ldots S_{i_n}^z > \; .
\ee
Here the angle brackets $<\ldots>$ imply statistical averaging with
a statistical operator $\hat\rho$. Expression (16) defines the
dipole-transition, or transverse, matrix. Equation (17) determines the
population-difference, or longitudinal, matrix. These are the matrices
with respect to the indices $i$ and $j$.

Spin density matrices [15] can be used for describing their entanglement
production under magnetic transitions [8]. They also appear when studying
the thermal entanglement caused by statistical operators of magnetic
systems [8,16,17]. However let us recall that in the present paper the
spin operators will be applied for describing not magnetic systems but
ensembles of radiating atoms. Similar pseudospin operators arise when
considering atomic entanglement in two-component Bose-Einstein condensates
[18] or in systems of atoms with several internal states [19,20].

The spatial variables $\br_i$ can be either discrete or continuous,
which is not essential, since the final results will not depend on this
difference. In the case of discrete $\br_i\in\Bbb{Z}^3$, one employs
summation over $i=1,2,\ldots,N$, while for continuous $\br_i\in\Bbb{R}^3$,
one should replace the sum $\sum_i$ by the integral $\int d\br_i$.

Single-partite spaces can be constructed on the basis of plane waves. For
the discrete spatial variable, one has
$$
\vp_k(\br_i) = \frac{1}{\sqrt{N}}\; e^{i\bk\cdot\br_i} \qquad
(\br_i\in\Bbb{Z}^3) \; ;
$$
while for the continuous variable,
$$
\vp_k(\br_i) = \frac{1}{\sqrt{V}}\; e^{i\bk\cdot\br_i} \qquad
(\br_i\in\Bbb{R}^3) \; ,
$$
$V$ being the system volume. Since the final results do not depend on
whether $\br_i$ is discrete or continuous, we shall employ the simpler
notation corresponding to the discrete case. The functions $\vp_k(\br_i)$
are orthogonal,
$$
\sum_k \vp_k^*(\br_i) \vp_k(\br_j) = \dlt_{ij} \; ,
$$
and form a complete basis, such that
$$
\sum_{i=1}^N \vp_k^*(\br_i) \vp_p(\br_i) = \dlt_{kp} \; .
$$

Denoting the vector
$$
|k>_i \; \equiv [\vp_k(\br_i)]
$$
as a column with respect to $i$, we define the single-partite space
$$
\cH_i \equiv \overline\cL\{|k>_i\}
$$
as a closed linear envelope over the basis $\{|k>_i\}$. Then the composite
space has the form $\cH=\otimes_i\cH_i$, in agreement with Eq. (1).

The first-order spin density matrix (16) is
\be
\label{18}
R_1(i,j) =\; < S_j^+ S_i^->\; .
\ee
For its trace, we have
\be
\label{19}
{\rm Tr}_{\cH_i}\; R_1 \equiv \sum_{i=1}^N R_1(i,i) =
\frac{N}{2}\; \left ( 1+ s \right ) \; ,
\ee
where the notation
\be
\label{20}
s \equiv \frac{2}{N} \sum_{i=1}^N < S_i^z>
\ee
is introduced.

The second-order spin density matrix (16) has the form
\be
\label{21}
R_2(i_1 i_2,j_1 j_2) = \;
< S_{j_2}^+ S_{j_1}^+ S_{i_1}^- S_{i_2}^- > \; .
\ee
Taking into consideration the properties of the spin operators, we see that
\be
\label{22}
\left [ ( 1  -\dlt_{i_1i_2})(1-\dlt_{j_1j_2}) -1 \right ]
R_2(i_1i_2,j_1j_2) = 0 \; .
\ee
Studying the semi-diagonal element
$$
R_2(il,jl) = R_2(li,lj) \; ,
$$
we shall invoke the decoupling
\be
\label{23}
< S_l^z S_j^+ S_i^-> \; = \; <S_l^z><S_j^+ S_i^-> \; ,
\ee
valid for $l\neq i,j$. Then we get
\be
\label{24}
R_2(li,lj) = (1 -\dlt_{il})(1-\dlt_{jl})\left (
\frac{1}{2}\; + < S_l^z>\right ) \; < S_j^+ S_i^-> \; .
\ee
This can be used for calculating the partial traces
\be
\label{25}
R_1^1 \equiv {\rm Tr}_{\cH_2} R_2 \; , \qquad
R_1^2 \equiv {\rm Tr}_{\cH_1} R_2
\ee
of $R_2$ defined on $\cH=\cH_1\otimes\cH_2$. The elements of $R_1^a$, with
$a=1,2$, are
$$
R_1^1(i,j) \equiv \sum_{l=1}^N R_2(il,jl) \; , \qquad
R_1^2(i,j) \equiv \sum_{l=1}^N R_2(li,lj) \; .
$$
The number of atoms $N$ is assumed to be large, $N\gg 1$. We find
\be
\label{26}
R_1^a(i,j) = \frac{N}{2}\; (1+s) R_1(i,j) \; .
\ee
From here, it is easy to get
\be
\label{27}
{\rm Tr}_\cH\; R_2 = \frac{N^2}{4} \; (1 + s)^2 \; .
\ee

The nonentangling counterpart of $R_2$ is proportional to $R_1^1\otimes
R_1^2$, with the proportionality constant obtained from the normalization
condition
$$
{\rm Tr}_\cH\; R_2 = {\rm Tr}_\cH \; R_2^\otimes \; .
$$
As a result, we come to
\be
\label{28}
R_2^\otimes = \frac{R_1^1\otimes R_1^2}{{\rm Tr}_\cH R_2} \; ,
\ee
where the equality
$$
{\rm Tr}_\cH \left ( R_1^1 \otimes R_1^2 \right ) =
\left ( {\rm Tr}_\cH\; R_2 \right )^2
$$
is used and ${\rm Tr}_\cH R_2$ is given by Eq. (27).

To find the entanglement produced by $R_2$, we need to define the restricted
norms
\be
\label{29}
||R_1||_\cD = \sup_k\; <k|R_1|k> \; , \qquad
||R_2||_\cD = \sup_{kp}\; <kp| R_2| pk> \; .
\ee
For this purpose, we shall use the single-mode laser approximation for the
correlator
\be
\label{30}
<S_i^+ S_j^->\; = \frac{w}{4}\; e^{-i\bk_0\cdot\br_{ij}} \qquad
(i\neq j) \; ,
\ee
where $\bk_0$ is a fixed wave vector of laser propagation and $\br_{ij}\equiv
\br_i-\br_j$. Then we obtain 
\be
\label{31}
||R_1||_\cD = \sup\left\{ \frac{1}{2}\; ( 1 + s), \; \frac{N}{4}\; w
\right \} \; .
\ee
For matrices (25) and (28), we find
\be
\label{32}
||R_1^a||_\cD = \frac{N}{2}\; (1 + s)||R_1||_\cD \; , \qquad
||R_2^\otimes||_\cD = ||R_1||^2_\cD \; .
\ee
The measure of entanglement produced by $R_2$ reads as
\be
\label{33}
\ep(R_2) = \log\; \frac{||R_2||_\cD}{||R_1||^2_\cD} \; .
\ee

To work out the norm $||R_2||_\cD$, we shall involve the decoupling
\be
\label{34}
<S_i^+ S_j^+ S_m^- S_n^->\; = \; <S_i^+ S_n^-><S_j^+ S_m^-> +
<S_i^+ S_m^-><S_j^+ S_n^-> \; ,
\ee
valid under condition that all indices are different. Then we get
\be
\label{35}
||R_2||_\cD = \sup\left\{ \frac{1}{2}\; (1+s)^2,\; \frac{N^2}{8}\;
w^2 \right \} \; .
\ee
Comparing Eqs. (31) and (32), we see that
\be
\label{36}
||R_2||_\cD = 2 ||R_1||^2_\cD \; .
\ee
Therefore norm (33) reduces to
\be
\label{37}
\ep(R_2) = \log 2 \; .
\ee
Thus, the entanglement produced by the transition spin density matrix $R_2$
is always constant.

Now, let us turn to the longitudinal spin density matrices (17). Such a
first-order matrix is
\be
\label{38}
Z_1(i,j) =\; <S_j^z S_i^z> \; .
\ee
As is evident,
\be
\label{39}
{\rm Tr}_{\cH_1} \; Z_1  = \sum_{i=1}^N Z_1(i,i) = \frac{N}{4} \; .
\ee

The second-order longitudinal spin density matrix writes as
\be
\label{40}
Z_2(i_1i_2,j_1j_2) = \; <S_{j_2}^z S_{j_1}^z S_{i_1}^z S_{i_2}^z > \; .
\ee
The matrix $Z_2$ is defined on $\cH=\cH_1\otimes\cH_2$. The partial traces give
the matrices
\be
\label{41}
Z_1^1 \equiv {\rm Tr}_{\cH_2}\; Z_2 \; , \qquad
Z_1^2 \equiv {\rm Tr}_{\cH_1} \; Z_2 \; ,
\ee
whose elements are
\be
\label{42}
Z_1^1(i,j) = \sum_{l=1}^N Z_2(il,jl) \; , \qquad
Z_1^2(i,j) = \sum_{l=1}^N Z_2(li,lj) \; .
\ee
We find
\be
\label{43}
Z_2(li,lj) =\frac{1}{4}\; < S_j^z S_i^z > \; , \qquad
Z_1^a(i,j) = \frac{N}{4}\; < S_j^z S_i^z> \; ,
\ee
for any $a=1,2$. From here,
\be
\label{44}
{\rm Tr}_\cH \; Z_2 = \frac{N^2}{16} \; .
\ee
The nonentangling counterpart of $Z_2$ is
\be
\label{45}
Z_2^\otimes  = \frac{Z_1^1\otimes Z_1^2}{{\rm Tr}_\cH Z_2} \; .
\ee

For the restricted norms, we have
\be
\label{46}
||Z_1||_\cD = \frac{1}{4}\; ( 1 + Ns^2 ) \; , \qquad
||Z_2^\otimes||_\cD = ||Z_1||^2_\cD \; .
\ee
The measure of entanglement produced by $Z_2$ is
\be
\label{47}
\ep(Z_2) = \log \; \frac{||Z_2||_\cD}{||Z_1||^2_\cD} \; .
\ee
Substituting to Eq. (47) the norm
\be
\label{48}
||Z_2||_\cD = \sup\left\{ \frac{3}{16}\; , \;
\frac{N^2}{16}\; s^4\right\} \; ,
\ee
we see that the entanglement production (47) strongly depends on the value
of $s$, which is the average population difference. When the latter is zero,
\be
\label{49}
\ep(Z_2) = \log 3 \qquad (s=0) \; .
\ee
But for any nonzero $s$, we have
\be
\label{50}
\ep(Z_2) = 0 \qquad (s\neq 0) \; ,
\ee
in view of large $N\gg 1$. The temporal behaviour of $s=s(t)$ depends on
particular physical systems.

\section{Collective Radiation}

Our aim is to find the temporal evolution of the average population
difference $s(t)$ under collective radiation of atoms. In this section,
the derivation of the pseudospin equations governing the behaviour of
$s(t)$ is given. This is done for resonant, effectively two-level, atoms
with dipole radiation transitions. One usually assumes that electrodipole
and magnetodipole transitions are described by equations possessing identical
structure. However there exists a difference between the equations for these
two types of transitions. In order to analyze this difference, we shall
consider here the case, when electrodipole transition is forbidden, and the
radiation occurs at magnetodipole transition.

The Hamiltonian of an atomic system, emitting electromagnetic radiation, is
\be
\label{51}
\hat H = \hat H_a + \hat H_f + \hat H_{af} \; .
\ee
The first term
\be
\label{52}
\hat H_a  =\sum_{i=1}^N \om_0 \left ( \frac{1}{2} + S_i^z \right )
\ee
corresponds to resonant atoms, with transition frequency $\om_0$. The field
Hamiltonian writes as
\be
\label{53}
\hat H_f = \frac{1}{8\pi} \int \left ( \bE^2 + \bH^2 \right ) \; d\br \; ,
\ee
where $\bE$ is the electric field and $\bH=\nabla\times\bA$ is the magnetic
field. The vector potential $\bA$ satisfies the Coulomb calibration
\be
\label{54}
\nabla \cdot \bA = 0 \; .
\ee
The last term $\hat H_{af}$ describes atom-field interactions [21].

When electric dipole transitions are forbidden, we need to consider other
radiation transitions. We study here the case of magnetic dipole transitions.
The matrix element of the magnetic dipole
\be
\label{55}
{\vec\mu}_{mn} \equiv \frac{1}{2c} \int \br \times \bj_{mn}(\br)\; d\br \; ,
\ee
with the light velocity, $c$, is expressed through the density of current
$$
\bj_{mn}(\br) = -\; \frac{ie}{2m_e}\; \left ( \psi_m^*\nabla\psi_n -
\psi_n\nabla\psi_m^* \right ) \; ,
$$
in which $e$ is electric charge, $m_e$ its mass, and $\psi_n=\psi_n(\br)$
are wave functions of states enumerated by the index $n=1,2$. Electric dipole
transitions are forbidden under conserved parity, when
\be
\label{56}
\psi_m^*(-\br)\psi_n(-\br) = \psi_m^*(\br)\psi_n(\br)
\ee
for all $m$ and $n$. Thence, the density of current is antisymmetric,
\be
\label{57}
\bj_{mn}(-\br) = -\bj_{mn}(\br) \; .
\ee
Denote the transition magnetic dipole
\be
\label{58}
\vec\mu \equiv\vec\mu_{21} \; , \qquad \vec\mu^* =\vec\mu_{12} \; .
\ee
Diagonal elements of dipole (55) are also nonzero, $\vec\mu_{nn}\neq 0$.
This is contrary to the case of electric dipole transitions, when parity
is not conserved, and only the nondiagonal dipole elements survive, with
diagonal ones being identically zero. The existence of nonzero diagonal
elements $\vec\mu_{nn}$ is in the basis of the differences between the
following equations for electric and magnetic dipole-transition radiation.

Under magnetic dipole transitions, the atom-field interaction Hamiltonian
writes as
\be
\label{59}
\hat H_{af} = - \sum_{i=1}^N {\bf M}_i \cdot {\bf B}_i +
\hat H_{af}' \; ,
\ee
where the operator of magnetic moment is
\be
\label{60}
{\bf M}_i \equiv \vec\mu \bS_i^+ + \vec\mu^* \bS_i^- \; ,
\ee
and the last term is
\be
\label{61}
\hat H_{af}'  = - \sum_{i=1}^N \left (\vec\mu_0 + \Dlt\vec\mu\; S_i^z
\right ) \cdot {\bf B}_i \; ,
\ee
with the notation
\be
\label{62}
\vec\mu_0 \equiv \frac{1}{2}\left ( \vec\mu_{11} + \vec\mu_{22}
\right ) \; , \qquad \Dlt\vec\mu \equiv \vec\mu_{22} -\vec\mu_{11} \; .
\ee
The total magnetic field is the sum
\be
\label{63}
{\bf B}_i = \bH_0 + \bH_i
\ee
of an external magnetic field $\bH_0$ and the radiation field
$\bH_i\equiv\bH(\br_i,t)$.

Using the commutation relations for the operators $\bE$, $\bA$ and $\bH$,
which can be found in book [22], we have the Heisenberg equations
\be
\label{64}
\frac{1}{c}\; \frac{\prt\bE}{\prt t} = \nabla\times\bH -\;
\frac{4\pi}{c}\; \bj \; , \qquad
\frac{1}{c}\; \frac{\prt\bA}{\prt t} = - \bE \; ,
\ee
in which the density of current is
\be
\label{65}
\bj = - c\sum_{i=1}^N \left ({\bf M}_i + \vec\mu_0 +\Dlt\vec\mu\; S_i^z
\right ) \times \nabla \dlt(\br -\br_i) \; .
\ee
The Heisenberg equations for the pseudospin operators are
\be
\label{66}
\frac{dS_i^-}{dt} = - i S_i^-\left ( \om_0 - \Dlt\vec\mu\cdot{\rm B}_i
\right )  - 2i S_i^z\vec\mu \cdot{\rm B}_i \; , \qquad
\frac{dS_i^z}{dt}  = i\left ( \vec\mu S_i^+ - \vec\mu^* S_i^- \right )
\cdot {\rm B}_i \; .
\ee

One usually considers the system of equations (64) and (66) for the field
and pseudospin operators. However, it is possible to eliminate the field
variables and to reduce the problem to the equations for only pseudospin
variables [21]. To this end, from Eqs. (64) we get the equation
\be
\label{67}
\left ( \nabla^2 - \; \frac{1}{c^2}\; \frac{\prt^2}{\prt t^2} \right )
\bA = - \; \frac{4\pi}{c}\; \bj \; ,
\ee
whose solution reads as
\be
\label{68}
\bA(\br,t) = \bA_{vac}(\br,t) + \frac{1}{c} \int
\bj\left (\br', t - \; \frac{|\br-\br'|}{c}\right ) \;
\frac{d\br'}{|\br-\br'|} \; ,
\ee
where $\bA_{vac}$ is the vacuum vector potential. Retardation in the dependence
of spin operators can be taken in the Born approximation
\be
\label{69}
S_j^-\left ( t -\; \frac{r}{c}\right ) \cong S_j^-(t) \Theta(ct-r)\;
e^{ik_0r} \; , \qquad
S_j^z\left ( t - \; \frac{r}{c} \right ) \cong S_j^z(t) \Theta(ct-r) \; ,
\ee
where $\Theta(\cdot)$ is the unit-step function.

Substituting the density of current (65) into potential (68), we have
\be
\label{70}
\bA_i = \bA_i^+ + \bA_i^- + \bA_i' +\bA_{vac} \; ,
\ee
where
$$
\bA_i^+(t) = -\sum_j \frac{1+ik_0 r_{ij}}{r_{ij}^3} \; \br_{ij} \times
\vec\mu\; S_j^+\left ( t -\; \frac{r_{ij}}{c}\right ) \; ,
$$
\be
\label{71}
\bA_i'(t) = -\sum_j \frac{1}{r_{ij}^3} \; \br_{ij} \times \left [ \vec\mu_0 
+ \Dlt\vec\mu\; S_j^z\left ( t  -\; \frac{r_{ij}}{c} \right ) \right ] \; ,
\ee
with the notation
$$
\br_{ij} \equiv \br_i - \br_j \; , \qquad r_{ij}\equiv|\br_{ij}| \; .
$$
Here the equality
$$
\frac{\prt}{\prt r}\; S_j^-\left ( t-\; \frac{r}{c}\right ) =
ik_0 S_j^- \left ( t-\; \frac{r}{c}\right )
$$
is taken into account, resulting from the Born approximation (69).

For the magnetic field
\be
\label{72}
\bH_i(t) \equiv \bH(\br_i,t) = \nabla_i\times \bA_i(t) \; ,
\ee
we obtain the expression
\be
\label{73}
\bH_i = \bH_i^+ + \bH_i^- + \bH_i' +\bH_{vac} \; ,
\ee
in which
$$
\bH_i^+(t) = \sum_j \left [ k_0^2 \;
\frac{\vec\mu-(\vec\mu\cdot{\bf n}_{ij}){\bf n}_{ij}}{r_{ij}} \; -
\left ( 1 + ik_0 r_{ij}\right ) \;
\frac{\vec\mu-3(\vec\mu\cdot{\bf n}_{ij}){\bf n}_{ij}}{r_{ij}^3}
\right ] \; S_j^+\left ( t-\; \frac{r_{ij}}{c}\right ) \; ,
$$
\be
\label{74}
\bH_i'(t) = -\; \sum_j
\frac{{\bf M}_j'-3({\bf M}_j'\cdot{\bf n}_{ij}){\bf n}_{ij}}{r_{ij}^3} \; ,
\ee
where ${\bf n}_{ij}\equiv\br_{ij}/r_{ij}$ and
\be
\label{75}
{\bf M}_j' \equiv \vec\mu_0 + \Dlt\vec\mu\; S_j^z\left ( t-\;
\frac{r_{ij}}{c}\right ) \; .
\ee
In what follows, we shall replace summation by integration according to the
rule
$$
\sum_{j=1}^N \Longleftrightarrow \rho \int_V d\br \; ,
$$
in which $\rho\equiv N/V$ is atomic density. Integrating over the spherical 
angle $\Om({\bf n})$, related to the unit vector ${\bf n}\equiv\br/r$, we 
shall  employ the properties 
$$
\int \left [ \vec\mu -(\vec\mu\cdot{\bf n}){\bf n} \right ] d\Om({\bf n})=
\frac{8\pi}{3}\; \vec\mu \; ,
$$
\be
\label{76}
\int \left [ \vec\mu -3(\vec\mu\cdot{\bf n}){\bf n} \right ] \;
d\Om({\bf n})= 0 \; , \qquad
\int (\vec\mu\cdot{\bf n}){\bf n} \; d\Om({\bf n}) =
\frac{4\pi}{3}\; \vec\mu \; .
\ee

In sums (71) and (74), there are the terms, where $j=i$ and $r_{ij}=0$, 
which yields divergence. These are the so-called self-action terms, whose 
divergence occurs because of the point-like representation of atoms. To 
avoid this divergence, the self-action must be treated more accurately, 
which can be done as follows.

Consider a single atom, located at $\br_j=0$. At the point $\br$, it creates
the vector potential
$$
\bA_s = \bA_s^+ +\bA_s^- +\bA_s' \; ,
$$
in which, according to Eqs. (71),
$$
\bA_s^+(\br,t) = -\; \frac{1+ik_0 r}{r^3} \; \br\times \vec\mu\; S^+
\left ( t -\; \frac{r}{c}\right ) \; , \qquad
\bA_s'(\br,t) = -\; \frac{1}{r^3}\; \br\times \left [ \vec\mu_0 +
\Dlt\vec\mu\; S^z\left ( t -\; \frac{r}{c}\right ) \right ] \; .
$$
The corresponding magnetic field
$$
\bH_s \equiv \nabla\times \bA_s = \bH_s^+ +\bH_s^- +\bH_s' \; ,
$$
contains, in agreement with Eqs. (74), the terms
$$
\bH_s^+(\br,t) = \left [ k_0^2 \;
\frac{\vec\mu-(\vec\mu\cdot{\bf n}){\bf n}}{r}\; - \left ( 1 + ik_0r
\right )\; \frac{\vec\mu-3(\vec\mu\cdot{\bf n}){\bf n}}{r^3} \right ]
S^+\left ( t -\; \frac{r}{c}\right ) \; ,
$$
$$
\bH_s'(\br,t) = -\;
\frac{{\bf M}'-3({\bf M}'\cdot{\bf n}){\bf n}}{r^3} \; ,
$$
where
$$
{\bf M}' \equiv \vec\mu_0 + \Dlt\vec\mu\; S^z\left ( t -\;
\frac{r}{c}\right ) \; .
$$

To consider the self-action, we need to go to the limit $\br\ra 0$.
Before doing this, let us average $\bH_s$ over spherical angles, which gives
$$
\bH_s \longrightarrow \frac{1}{4\pi} \; \int \bH_s(\br,t) \; d\Om(\br) =
\frac{2k_0^2}{3r}\left [ \vec\mu\; S^+\left ( t-\; \frac{r}{c}\right ) +
\vec\mu^* S^-\left ( t-\; \frac{r}{c}\right ) \right ] \; .
$$
Keeping in mind $r\ra 0$, we may present the exponential $e^{ik_0r}$ as an
expansion
$$
e^{ik_0r} \simeq 1 + ik_0 r \qquad (r\ra 0) \; .
$$
Imitating the atom nonlocality, we may average the singular term $1/r$ over
the distance $r$ between the electron wavelength $\lbd_e\equiv2\pi\hbar/m_ec$
and the radiation wavelength $\lbd_0\equiv 2\pi c/\om_0$, so that
$$
\frac{1}{\lbd_0-\lbd_e} \; \int_{\lbd_e}^{\lbd_0} \frac{dr}{r} =
\frac{k_0}{2\pi} \; \ln\left ( \frac{m_ec^2}{\hbar\om_0}\right ) \; ,
$$
where we take into account that $\lbd_e\ll\lbd_0$. Then, for the averaged
in this way self-action magnetic field, we find
$$
\bH_s(t) = \frac{2}{3}\; ik_0^3\; \left [ \vec\mu^* S^-(t) -\vec\mu\; S^+(t)
\right ] + \frac{k_0^3}{3\pi} \; \ln\left ( \frac{m_ec^2}{\hbar\om_0}
\right ) \left [ \vec\mu^* S^-(t) +\vec\mu\; S^+(t) \right ] \; .
$$

Introduce the notation for the natural width
\be
\label{77}
\gm_0 \equiv \frac{2}{3}\; |\vec\mu|^2 k_0^3
\ee
and the Lamb frequency shift
\be
\label{78}
\dlt_L \equiv \frac{\gm_0}{2\pi} \; \ln\left ( \frac{m_ec^2}{\hbar\om_0}
\right ) \; .
\ee
Equations (66), for a single atom with no external fields, are transformed into
the system of equations
$$
\frac{dS^-}{dt} = - i (\om_0 - \dlt_L -i\gm_0) S^- -\;
\frac{\vec\mu^2}{|\vec\mu|^2}\; \left ( \gm_0 + i\dlt_L \right ) S^+ +
\frac{\Dlt\vec\mu\cdot\vec\mu}{|\vec\mu|^2}\; (\gm_0 + i\dlt_L)
\left ( \frac{1}{2}\; - S^z\right ) \; ,
$$
\be
\label{79}
\frac{dS^z}{dt} = - 2\gm_0 \left ( S^z +\frac{1}{2} \right ) \; .
\ee
These equations show that the influence of the radiation self-action reduces
to the appearance of the longitudinal attenuation $\gm_1=2\gm_0$ and transverse
attenuation $\gm_2=\gm_0$. When the atom is immersed into a medium, one usually
treats $\gm_1$ and $\gm_2$ as independent parameters, which is assumed in what
follows. The Lamb shift can be included in the definition of the transition
frequency $\om_0$.

Thus, the influence of self-action will be taken into consideration by
incorporating the related attenuation terms with $\gm_1$ and $\gm_2$ into
the evolution equations. And in the magnetic field $\bH=\nabla\times\bA$, the
self-action will be excluded. Then the field $\bH=\bH(\br,t)$, in view of Eq.
(72), can be rearranged to
\be
\label{80}
\bH =\bH_{non} + \bH_{dip} + \bH_{vac} \; ,
\ee
where the nondipole part
\be
\label{81}
\bH_{non} = \bH^+_{non} + \bH_{non}^-
\ee
is due to the field created by radiating atoms, without the dipole part
\be
\label{82}
\bH_{dip} = \bH^+_{dip} + \bH^-_{dip} + \bH' \; .
\ee
In the corresponding expressions for the nondipole field,
\be
\label{83}
\bH_{non}^+(\br_i,t) = \sum_{j(\neq i)} k_0^2\;
\frac{\vec\mu-(\vec\mu\cdot{\bf n}_{ij}){\bf n}_{ij}}{r_{ij}} \;
S^+_j\left ( t -\; \frac{r_{ij}}{c}\right )
\ee
and for the dipole part
\be
\label{84}
\bH_{dip}^+(\br_i,t) = - \sum_{j(\neq i)} (1+ik_0 r_{ij})\;
\frac{\vec\mu-3(\vec\mu\cdot{\bf n}_{ij}){\bf n}_{ij}}{r^3_{ij}} \;
S^+_j\left ( t -\; \frac{r_{ij}}{c}\right )
\ee
the self-action terms are omitted. The last term in Eq. (82), $\bH'(\br,t)$,
for $\br=\br_i$, is given by $\bH_i'(t)\equiv\bH'(\br_i,t)$ from Eq. (74),
again with no self-action terms.

Introduce the radiating symmetric part of the total radiation field,
\be
\label{85}
\bH_{rad} \equiv \frac{1}{4\pi} \int \left ( \bH_{non} +\bH_{dip}
\right ) \; d\Om(\br) = \frac{1}{4\pi} \int \bH_{non}(\br,t) \;
d\Om(\br) \; .
\ee
This can be decomposed as
\be
\label{86}
\bH_{rad} = \bH^+_{rad} + \bH_{rad}^- \; ,
\ee
with
\be
\label{87}
\bH^+_{rad}(\br_i,t)  =\frac{2}{3}\; k_0^2\vec\mu \; \sum_{j(\neq i)}
\frac{1}{r_{ij}} \; S_j^+\left ( t -\; \frac{r_{ij}}{c}\right ) \; .
\ee
Hence, the total radiation field can be represented as
\be
\label{88}
\bH =\bH_{rad} + \Dlt\bH + \bH_{vac} \; ,
\ee
with the notation
\be
\label{89}
\Dlt\bH \equiv \bH_{non} + \bH_{dip} - \bH_{rad} \; .
\ee

The anisotropic part of the radiation field contributes to the expression
\be
\label{90}
\xi(\br,t) \equiv -2i \vec\mu \cdot \left ( \bH_{vac} + \bH_{non} +
\bH_{dip} - \bH_{rad} \right ) \; ,
\ee
for which
$$
\int \xi(\br,t) \; d\Om(\br) = 0 \; .
$$
Equation (90) describes local field fluctuations. We also define
\be
\label{91}
\xi_0(\br,t) \equiv -\Dlt\vec\mu \cdot \bH(\br,t) \; ,
\ee
where $\bH$ is given by Eq. (88), and the frequency shift
\be
\label{92}
\Dlt_0 \equiv -\Dlt\vec\mu \cdot \bH_0 \; .
\ee

Then we consider the statistical averaging of the pseudospin equations 
(66) in the frame of the scale-separation approach [21]. In this course, 
we define the statistical averages for the {\it transition function}
\be
\label{93}
u(\br,t) \equiv 2<S^-(\br,t)>\; ,
\ee
{\it coherence intensity}
\be
\label{94}
w(\br,t) \equiv 4 <S^+(\br,t) S^-(\br+0,t> \; ,
\ee
and {\it population difference}
\be
\label{95}
s(\br,t) \equiv 2<S^z(\br,t)> \; ,
\ee
in which the angle brackets $<\ldots>$ imply the averaging over the pseudospin
degrees of freedom, not involving the fluctuating fields (90) and (91) treated
as random variables. Function (94) is to be understood as the limit
$$
w(\br,t) = 4\lim_{|\br-\br'|\ra+0}\; < S^+(\br,t) S^-(\br',t)> \; .
$$
In the mean-field approximation, the latter simplifies to $w=|u|^2$. Definitions
(93), (94), and (95) are in agreement with Eqs. (20) and (30).

Also, introduce the effective field acting on atoms as
\be
\label{96}
f = f_0 + f_{rad} + \xi \; ,
\ee
in which
\be
\label{97}
f_0 \equiv -2i\vec\mu \cdot \bH_0
\ee
is due to an external magnetic field; the term
\be
\label{98}
f_{rad} \equiv -2i\; <\vec\mu \cdot\bH_{rad}>
\ee
is caused by the isotropic radiation field (85); and $\xi$ is the local fluctuating
field (90). Allowing for Eqs. (86) and (87), field (98) takes the form
\be
\label{99}
f_{rad}(\br,t) = - i\gm_0\rho \int\left [ G(\br-\br',t) u(\br',t) +
\frac{\vec\mu^2}{|\vec\mu|^2}\; G^*(\br-\br',t) u^*(\br',t)\right ] \; d\br' \; ,
\ee
with the transfer function
$$
G(\br,t) \equiv \frac{\exp(ik_0r)}{k_0r} \; \Theta(ct-r) \; .
$$

Finally, from Eqs. (66) we derive the evolution equations
$$
\frac{\prt u}{\prt t} = -i(\om_0 +\Dlt_0 +\xi_0 -i\gm_2) u + fs \; ,
$$
$$
\frac{\prt w}{\prt t} = - 2\gm_2 w + \left ( u^* f + f^* u
\right ) s\; , 
$$
\be
\label{100}
\frac{\prt s}{\prt t} = -\; \frac{1}{2}\left ( u^* f + f^* u\right ) -
\gm_1(s-\zeta) \; ,
\ee
where $\zeta$ is a stationary population difference of an atom. These
equations differ from analogous equations for electric dipole transitions
[21,23] by the presence in Eqs. (100) of the frequency shifts $\Dlt_0$
and $\xi_0$. Both these shifts are due to $\Dlt\vec\mu$, given in Eqs. (62),
which is nonzero because of the nonvanishing diagonal elements $\vec\mu_{nn}$.
The constant part of the summary shift $\Dlt_0+\xi_0$ can be neglected as
compared to $\om_0$. And the fluctuating part of this shift results in the
appearance of an additional inhomogeneous broadening $\gm_2^*$, so that
the total line width becomes the sum $\gm_2+\gm_2^*$.

\section{Entanglement by Superradiance}

According to Sec. 3, the entanglement-production measure $\ep(Z_2)$,
caused by the spin density matrix $Z_2$, defined in Eq. (17), strongly
depends on the value $s=s(t)$ of the population difference. Changing $s(t)$
would govern the level of produced entanglement. To make such an evolutional
entanglement efficient, one needs that $s(t)$ would cross the value $s=0$,
when the measure $\ep(Z_2)$ jumps between expressions (49) and (50). Thus,
we have to analyze the temporal behaviour of the population difference
$s(t)$, which is described by Eqs. (100).

It seems that an adequate method of quickly changing the population difference
between $s\neq 0$ and $s=0$ could be by means of the process of superradiance.
Then, at the initial time $t=0$, one prepares the atomic system in the inverted
state, with $s\approx 1$. The superradiance burst is peaked at the delay time
$t_0$, when $s\approx 0$, after which $s(t)$ goes rapidly to $s=-1$.

Temporal dynamics of superradiance, described by Eqs. (100), has been
investigated for atomic systems [21,23,24] as well as for spin superradiance
[21,25,26]. Therefore, we shall not repeat here the solution of Eqs. (100) but
will straightforwardly go to results.

Let the atomic system at the initial time be inverted, with $s_0\equiv s(0)>0$.
The effective coupling parameter of atomic interactions through radiation field
is
\be
\label{101}
g \equiv \rho \; \frac{\gm_0}{\gm_2} \; \int \frac{\sin(k_0r-kz)}{k_0r}\;
d\br \; ,
\ee
where $k\equiv\om/c$ is the wave vector of the seed field selecting the
longitudinal propagating mode. The frequency $\om$ of the seed field is in
resonance with the atomic transition frequency $\om_0$. Because of $\om\approx
\om_0$, one has $k\approx k_0$. Substantial coherence develops in the system
if the coupling parameter (101) is large, such that $gs_0\gg 1$.

Consider a purely self-organized process, when no coherence is imposed upon
the system at $t=0$, so that $w_0\equiv w(0)=0$. Then the relaxation process
starts with the incoherent quantum stage of spontaneous radiation, which lasts
till the crossover time
\be
\label{102}
t_c = \frac{T_2}{2gs_0} \; ,
\ee
where $T_2\equiv 1/\gm_2$. For $gs_0\gg 1$, the crossover time is small,
$t_c\ll T_2$. After time (102), coherent effects become important and a
superradiance pulse arises, with the intensity of radiation peaking at the
delay time
\be
\label{103}
t_0 =  t_c \left ( 1 +\ln\left | \frac{2}{\gm_3 t_c}\right | \right ) \; .
\ee
Here $\gm_3$ is the dynamic inhomogeneous broadening defined as
$$
\gm_3 \equiv {\rm Re}\; \lim_{\tau\ra\infty} \; \frac{1}{\tau} \;
\int_0^\tau dt \; \int_0^t \ll \xi^*(t)\xi(t')\gg \; \exp\left\{ -
(i\om_0 +\gm_2 - \gm_2 gs)(t-t')\right \} \; dt' \; ,
$$
where $\ll\ldots\gg$ implies the stochastic averaging over the random
fluctuating fields (90). At the delay time (103), the value $\gm_3 t_c$,
approximately, is
$$
\gm_3 t_c \approx \frac{1}{2gs_0} \ll 1 \; ,
$$
because of which $t_0>t_c$. The superradiant burst is rather narrow, with the
pulse time
\be
\label{104}
\tau_p = \frac{T_2}{gs_0} \; .
\ee
At the transient coherence stage, when $t_c<t\ll T_1$, where $T_1\equiv
1/\gm_1$, the population difference behaves as
\be
\label{105}
s =\frac{1}{g}\; - s_0{\rm tanh}\left ( \frac{t-t_0}{\tau_p} \right ) \; .
\ee
The first term here is small, provided that $g\gg 1$. After the time $t_0$,
function (105) rapidly diminishes to the value about $s=-s_0$. The population
difference passes through $s=0$ at the time, which is very close to $t_0$.
Thus, the superradiant regime makes it possible to realize a very rapid
passage of the population difference through zero.

Then one can invert the system again by a $\pi$-pulse. In the process
of this inversion, $s$ also crosses zero. Upon regaining inversion, a
new superradiant regime develops, and $s$ crosses zero after the delay
time $t_0$, counted from the moment of the novel inversion. This procedure
can be repeated many times. Each time, when $s=0$, entanglement production
reaches its maximum, with the measure $\ep(Z_2)$ jumping from zero to $\log
3$, in accordance with Eqs. (49) and (50). These jumps are so sharp because
of the large number of atoms $N\gg 1$ and, respectively, owing to well
developed coherent effects. A series of superradiant bursts, generated
by the repeated preparation of an inverted atomic system, can be called
punctuated superradiance. Such a regime can also be realized for the case
of spin systems, resulting in punctuated spin superradiance [26,27]. This
regime can be employed for regulating the related entanglement production
and for generating a punctuated series of sharp signals of produced
entanglement. The described regime allowing for the generation of a
regulated series of sharp entanglement pulses can be called the {\it
punctuated entanglement production}.

The nontrivial behaviour of evolutional entanglement, considered in
this paper, is due to the entanglement production $\ep(Z_2)$ caused by the
second-order spin density matrix $Z_2$. Generally speaking, we could study
as well the entanglement produced by higher-order spin density matrices
$Z_n$. However, emphasis has been placed on $Z_2$ because of the special
role of second-order density matrices in statistical mechanics [28] and
for quantum information processing [29,30].

\vskip 5mm

{\bf Acknowledgement}

\vskip 2mm

I am grateful to E.P. Yukalova for discussions and advice.

\end{document}